\documentstyle[aps]{revtex}
\input{epsf.tex}
\begin{document}
\draft
\title{Collective mode of homogeneous superfluid Fermi gases in the BEC-BCS crossover}
\author{R. Combescot$^{\,a}$, M.Yu. Kagan$^{\,b}$ and S. Stringari$^{\,c}$}
\address{(a) Laboratoire de Physique Statistique,
 Ecole Normale Sup\'erieure,
24 rue Lhomond, 75231 Paris Cedex 05, France \\
and Institut Universitaire de France}
\address{(b) P.L. Kapitza Institute
    for Physical Problems, Kosygin street 2, Moscow, Russia, 119334}
\address{(c) Dipartimento di Fisica, Universit\`a di Trento and BEC-INFM, I-38050 Povo, Italy}
\date{Received \today}
\maketitle
\pacs{PACS numbers : 05.30.Fk,  32.80.Pj, 47.37.+q, 67.40.Hf }

\begin{abstract}
We perform a detailed study of the collective mode across the whole BEC-BCS crossover in fermionic gases at zero temperature, covering the whole range of energy beyond the linear regime. This is done on the basis of the dynamical BCS model. We recover first the results of the linear regime in a simple form. Then specific attention is payed to the non linear part of the dispersion relation and its interplay with the continuum of single fermionic excitations. In particular we consider in detail the merging of collective mode into the continuum of single fermionic excitations. This occurs not only on the BCS side of the crossover, but also slightly beyond unitarity on the BEC side. Another remarkable feature is the very linear behaviour of the dispersion relation in the vicinity of unitarity almost up to merging with the continuum. Finally, while on the BEC side the mode is quite analogous to the Bogoliubov mode,  a difference appear at high wavevectors. On the basis of our results we determine the Landau critical velocity in the BEC-BCS crossover which is found to be largest close to unitarity. Our investigation has revealed interesting qualitative features which would deserve experimental exploration as well as further theoretical studies by more sophisticated means.
\end{abstract}

\section{INTRODUCTION}

The field of ultracold fermionic atoms has seen recently a remarkable achievement. With $^6$Li as well as with $^{40}$K, it has proved to be possible to display an experimental realization of a system going continuously from a Bose-Einstein superfluid to a BCS one \cite{levico}. This has been done by taking advantage of a Feshbach resonance, which allows to control by a magnetic field the scattering length, and thereby the effective interaction, for two fermionic atoms belonging to different hyperfine states (hereafter referred to as spin up and spin down states, as it is commonly done by analogy with electrons forming pairs in a superconductor). On one side of the resonance the scattering length $a$ is positive and the spin up and spin down atoms can form diatomic molecules. At low enough temperature these bosonic molecules can undergo a Bose-Einstein condensation which has indeed been observed for $^{40}$K \cite{expK40} as well as for $^6$Li \cite{expLi1,expLi2,Bourdel}. On the other side of the resonance where the scattering length is negative, evidence for BCS superfluidity has also been observed, for example through pair-breaking features in collective modes \cite{collmode}, observation of the pairing gap \cite{Chin} and more directly and recently observation of vortices \cite{Zwierlein}. When the magnetic field is tuned across the resonance, where the scattering length diverges, the superfluid undergoes a crossover from Bose-Einstein condensates to a BCS superfluid. Around resonance the superfluid is, so to speak, of a new kind since it is neither a full BCS superfluid nor a full molecular condensate. It is obviously quite interesting to study in detail this new superfluid, although it is clear that it will share the properties common to a BEC and a BCS superfluid.

In this paper we will be interested specifically by the way in which the Bogoliubov collective mode, which is well known for Bose condensates since it is the only elementary excitation allowed in this system, goes over into the so-called Anderson-Bogoliubov mode which is known to exist in the BCS case. In the limit of very low frequency this is easy to understand since in both limits this mode is just identical physically to sound waves, propagating in the neutral superfluid, which are coupled to fluctuations of the phase of the order parameter. Naturally the same interpretation will also hold in the crossover. However we will be interested in this paper with all the range of possible frequencies, which will clearly display a larger number of interesting features. In particular it is worthwhile to enquire how the composite nature of the boson corresponding to the mode will show up in the dispersion relation.

Our study of the collective mode will rely on the use of dynamical BCS theory. Naturally this theory is not exact in the crossover, while it is known for quite a long time to give a correct description not only in the BCS limit, but also in the deep BEC regime \cite{note1}. In between it should be regarded as an interpolation model, which has no specific reason to be quantitatively reliable. On the other hand BCS theory is a very coherent framework which is known to display all the qualitative physics on both sides of the crossover and it is therefore reasonable to trust it qualitatively in the crossover, and to make use of it at least as a first instructive step. Actually it has already been used in the literature for some partial studies of the collective mode. For example Minguzzi, Ferrari and Castin \cite{mfc}, calculating the dynamic structure factor, have stressed that the observation of this mode in a trapped ultracold gas would be a clear signature of superfluidity. BCS theory has also been used by B\"uchler, Zoller and Zwerger \cite{bzz} to consider essentially the low frequency regime. Finally Pieri, Pisani and Strinati \cite{pps} had also to consider the collective mode, in the course of their calculation of the spectral function which was their major focus and which was mostly aimed at the physics of high $T_c$ superconductors, but they did not proceed to a detailed study as we will do in the present paper.

The paper is organized as follows. In the next section, as an introduction, we recall for the static properties the standard BCS model widely used to describe the BEC-BCS crossover. We take advantage of it to introduce some new results concerning the sound velocity and also some simple relations which appear within this model around the $\mu =0$ point. Then we give the principles behind the proper handling of the BCS dynamics, the technical details being deferred to the appendix. We present the result for the collective mode dispersion relation and we give our results for the detailed study of this mode, on the BCS side, on the BEC side and around unitarity. Once again, for clarity and simplicity, the technical analytical details related to this study are given only in the appendix.

\section{STATIC BCS FRAMEWORK}
\label{mf}

Let us first briefly recall the standard BCS framework and how it is applied to the BEC-BCS
crossover, in the way initiated by Leggett \cite{legg} and by Nozi\`eres and Schmitt-Rink \cite{nsr}, and developped in particular by S\' a de Melo, Randeria and Engelbrecht \cite{sdm,gss}. This BCS model has been for example used
recently \cite{vgps} by Viverit \emph{et al}, together with a semiclassical approximation, to calculate the momentum distribution in trapped
gases across the BEC-BCS crossover. This is often called the mean-field BCS model. However there is no diagonal mean-field included in this framework \cite{note2}, the only mean-field character being just the fact that BCS theory itself is formally a mean-field theory. For this reason we drop the "mean-field" qualificative since it is redundant at best, and misleading at worst.

In the BCS framework one eliminates
the usual high energy cut-off $\omega _c$ and the attractive interaction $V$, in the standard gap equation:
\begin{eqnarray}\label{eqga}
\frac{1}{V}=  \sum_{k}^{\omega _c} \frac{1}{2E_k}
\end{eqnarray}
in favor of the scattering length $a$, which leads to:
\begin{eqnarray}\label{eqgap}
\frac{m}{4 \pi a} =  \sum_{k} \left( \frac{1}{2\epsilon _k}-\frac{1}{2E_k}\right)
\end{eqnarray}
where $\epsilon _k = k^2/2m$ is the atomic kinetic energy and $ E_k = \sqrt{\xi _{k}^{2}+\Delta ^{2}}$ the single particle excitation energy. Here we have set $\xi_k=\epsilon _k -\mu $, with $\mu $ being the chemical potential and $\Delta $ the
gap parameter, and we take $\hbar=1$ throughout the paper. This gap equation is supplemented by the
equation giving the atom density $n$ for a single spin state:
\begin{eqnarray}\label{eqpartnumb}
n \equiv \frac{k_{F}^{3}}{6\pi ^2} =  \sum_{k} \frac{1}{2} \left(1-\frac{\xi_k}{E_k}\right)
\end{eqnarray}
This equation, which does not play any role in the standard BCS weak coupling theory, is here
an essential ingredient. Together with Eq.(\ref{eqgap}), it gives the evolution of the gap and the
chemical potential, as a function of $1/k_Fa$, across the BEC-BCS crossover. After some manipulations we rewrite Eq.(\ref{eqpartnumb}) as:
\begin{eqnarray}\label{eqk0}
k_F^{3} =\frac{{\Delta}^{2}}{2m}\; J_4 \hspace{10mm} J_4 =  \int_{0}^{\infty}\!d{ k} \;\;\frac{{ k}^4}{E_{k}{^3}}
\end{eqnarray}
Correspondingly we can rewrite Eq.(\ref{eqgap}) as:
\begin{eqnarray}\label{eqk0a}
\frac{\pi }{2 a} =\frac{{\Delta}^2+\mu ^{2}}{m}\;J_2 - \frac{\mu }{2\,m^2}\;J_4\hspace{10mm}J_2 =  \int_{0}^{\infty}\!d{k} \;\;\frac{{k}^2}{E_{k}^3}
\end{eqnarray}

In the BEC limiting case where $\mu \rightarrow - \infty$ and $\Delta / \mu  \rightarrow 0$ one has $J_2=(\pi /16).(2m/|\mu |)^{3/2}$ and $J_4=(3\pi /16).(2m)^{5/2}/|\mu |^{1/2}$. One finds naturally from Eqs.(\ref{eqgap}-\ref{eqpartnumb})  $|\mu |=1/(2ma^2)$, and 
$\Delta /|\mu | = (32\pi na^3)^{1/2}$.

\subsection{Vicinity of $\mu =0$}
It is of interest to consider more closely the vicinity of $\mu =0$, which is obtained for $1/(k_Fa)_0 \simeq 0.553$, since it is around this region that the gas will switch from a typical BCS behaviour to a typical Bose one. We consider the results of first order expansion in the vinicity of $\mu =0$. Details are given in the Appendix. It turns out that all the specific numerical integrals coming in this calculation can be expressed in terms of Euler's gamma function $\Gamma (1/4)$. This leads to very simple relations between various physical quantities in terms of the universal parameter $1/(k_Fa)_0$ or equivalently $(\Delta/E_F)_{0}$. One finds indeed that the gap to Fermi energy ratio satisfies merely:
\begin{eqnarray}\label{}
\left(\frac{\Delta}{E_F}\right)_{0}^{2} = 2 \left(\frac{1}{k_Fa}\right)_{0} 
\end{eqnarray}
where the index $0$ indicates that the relevant physical quantities are taken for $\mu =0$. The behaviour of $\Delta/E_F$ and $1/k_Fa$ in the vicinity of $\mu =0$, can   be simply expressed in terms of $(\Delta/E_F)_0$ or, equivalently $(1/k_Fa)_0$.
One finds:
\begin{eqnarray}\label{exp1}
\frac{\Delta}{E_F}= \left(\frac{\Delta}{E_F}\right)_{0}
 \left( 1 - \frac{\pi }{4} \left(\frac{1}{k_Fa}\right)_{0} \frac{\mu }{E_F}\right)
\end{eqnarray}
and
\begin{eqnarray}\label{exp2}
\frac{1}{k_Fa} = \left(\frac{1}{k_Fa}\right)_{0}
 \left( 1 - \left[\frac{\pi }{8} \left(\frac{1}{k_Fa}\right)_{0} +
 \frac{1}{\pi } \left(k_Fa\right)^2_{0}\right] \frac{\mu }{E_F}\right) \; .
\end{eqnarray}
It would be  interesting to compare these results with the ones of Monte-Carlo calculations and with experiments to check how good is the quantitative description of BCS theory in this respect.

\subsection{Sound velocity}\label{vs}
Finally let us consider in this model the sound velocity $c_s$, given by $c_{s}^{2}=(n/m) (\partial \mu / \partial n)$. Varying Eq.(\ref{eqgap}) gives $\partial \mu /\partial \Delta =  J_2 {\Delta} /J_{\xi}$, where we
have introduced:
\begin{eqnarray}\label{eqjc}
J_{\xi} = \frac{1}{2m} J_4 - \mu  J_2=\int_{0}^{\infty}\!d{k} \;\;\frac{{k}^2\,{\xi}_{k}} {E _{k}^3} = m \int_{0}^{\infty}\!d{k} \;\;\frac{1} {E_k}
\end{eqnarray}
while varying Eq.(\ref{eqpartnumb}) leads to $4\pi ^2 (\partial n/\partial \mu)_{\Delta} = J_2 {\Delta}^2$ and $4\pi ^2 (\partial n/\partial \Delta)_{\mu }   =  J_{\xi} {\Delta}$. Taken together with Eq.(\ref{eqk0}) this leads to:
\begin{eqnarray}\label{eqcs}
c_{s}^{2}= \frac{1}{3m^2}\; \frac{J_2 J_4 {\Delta}^2}{J_{2}^{2} {\Delta}^2 + J_{\xi}^{2}} 
\end{eqnarray}
In the BCS limit, where $\mu \simeq E_F \equiv k_{F}^{2}/2m  > 0 $ and ${\Delta}/\mu  \rightarrow 0_{+}$, one finds  from Eq.(\ref{eqk0}) $J_4=k_{F}^{2} J_2=2m k_{F}^{3}/{\Delta}^2$ while $J_{\xi} \sim - (2 m^2/k_F) \ln (\Delta/E_F)$ is negligible due to particle-hole symmetry, leading to the well known result $c_{s}^{2}= k_{F}^{2}/3m^2$. On the other hand the dependence of $n$ on $\Delta $, i.e. $J_{\xi}$, can not be omitted in the rest of the BEC-BCS transition. For example at unitarity, Eq.(\ref{eqk0a}) implies $J_4=2mJ_2 ({\Delta}^2+\mu ^{2})/\mu$, while accordingly Eq.(\ref{eqjc}) gives $J_{\xi} = J_2 {\Delta}^2/\mu $. When this is inserted in Eq.(\ref{eqcs}) this merely leads to $c_{s}^{2}= 2 \mu /(3m)$. But this is what is expected at unitarity from universality arguments proving that, for dimensional reasons, $\mu $ is proportional to $k_{F}^{2}$, i.e. $n^{2/3}$, which give $c_{s}^{2}=(n/m)(\partial \mu / \partial n)=(2/3) \mu /m $. Hence, at unitarity, BCS theory leads naturally to a result in agreement with universality, but only provided that the important dependence of $n$ and $\mu $ on $\Delta $ is retained.

This is also the case in the BEC limit, where ${\Delta}/\mu  \rightarrow 0_{-}$. Indeed we have seen that $J_4=6m|\mu | J_2$ and $J_{\xi}=4 |\mu | J_2$,
leading to $c_{s}^{2}=\Delta^2/(8m|\mu |)=2\pi na / m^2$. This coincides with the standard sound velocity $c_{s}^{2}=g_M n_M/m_M$ of a molecular BEC gas with $n_M
=n$, $m_M=2m$ and $g_M=4\pi a_M/m_M$ where $a_M=2\,a$ is the value for the molecular scattering length predicted by the BCS model \cite{sdm}, as opposed to the exact result \cite{petrov,bkkcl} $a_M=0.6\,a$. 

Finally in the case $\mu =0$ we end up with:
\begin{eqnarray}\label{}
c_{s}^{2}=\frac{k_{F}^{2}}{3m^2}\; \frac{(\pi/2)(1/k_Fa)_{0}^{2}}{1+(\pi^2/8)(1/k_Fa)_{0}^{3}} \simeq 0.132\, \frac{k_{F}^{2}}{m^2}
\end{eqnarray}
which can also be obtained from the relationship  $c_{s}^{2}=(n/m) (\partial \mu/ \partial n)$ with $\mu(n)$ given by the expansion Eq.(\ref{exp2}).

\section{BCS DYNAMICS}

In order to obtain the collective mode, we have now to deal with the dynamics of the BCS theory. This can be done following various approaches, which are actually all basically equivalent presentations of the standard self-consistent BCS theory. To be specific we will consider here the kinetic equation approach. More details on this method will be given in the Appendix, where we will also sketch the derivation following the Green's function, or diagrammatic, approach to the BCS theory.

The important physical point about this dynamics is that it has to be self-consistent. This means that not only the single particle excitations are set in motion, but the order parameter has also space and time dependences. This already appears clearly from the results given in the preceding section for sound velocity where, in order to obtain the proper result, one has to consider $(\partial n/\partial \Delta)_{\mu }$, in addition to $(\partial n/\partial \mu)_{\Delta}$. This shows that one has to consider fluctuations of the modulus $|\Delta |$ of the order parameter. Similarly one has also to consider fluctuations of the phase of the order parameter, since its gradient is directly linked to the superfluid velocity. Hence the proper physical description is the dynamical evolution of the excitations in the presence of a dynamical order parameter, which is itself self-consistently determined by the excitations. If we did not take these fluctuations of the order parameter into account, we would have \cite{schr} a dynamics which would violate gauge invariance and particle conservation, and at the same time would not display the gapless collective mode. It is well known \cite{schr} that the existence of the gapless collective mode is intimately related to gauge invariance.

These features are quite clear in the kinetic equation formulation of the BCS theory, which is a generalization to the superfluid case of the standard Landau's kinetic equation for a Fermi liquid.
In addition to the diagonal self-consistent fields present in Landau theory, superfluidity introduces off-diagonal self-consistent fields and leads to deal with the following 2x2 matrix for the single particle density:
\begin{eqnarray}
({\hat n}_{k\,k'})_{\lambda\mu }=\langle c_{k\,\lambda }^{\dag}c_{k'\,\mu } \rangle 
\end{eqnarray} 
where $\lambda,\mu = 1$ or $2$, and by definition $c_{k,1}=c_{k\,\uparrow}$ and
$c_{k,2}=c^{\dag}_{-k\,\downarrow}$ are Nambu doublets. 
At the level of small fluctuations with wavevector $q$, the fluctuating part $\delta {\hat n}_k$ of this density matrix satisfies the kinetic equation:
\begin{eqnarray}\label{eqkin}
\omega \delta {\hat n}_k =  \delta {\hat n}_k  \;{\hat \epsilon} ^{0}_{k+q/2} - {\hat \epsilon} ^{0}_{k-q/2} \; \delta {\hat n}_k
+ {\hat n}^{0}_{k-q/2} \;\delta {\hat \epsilon}_k  - \delta {\hat \epsilon}_k  \;{\hat n}^{0}_{k+q/2}
\end{eqnarray}
which is formally identical to the standard kinetic equation, except that one deals now with 2x2 matrices.
The equilibrium energy matrix ${\hat \epsilon}^{0}_{k}$, together with its fluctuating part $\delta {\hat \epsilon}_k$, and the corresponding equilibrium density matrix ${\hat n}^{0}_{k}$ are given explicitly in the appendix.

As usual we find the dispersion relation for the collective mode by looking at the momentum and frequency $({\bf q},\omega )$ which makes the response of the system infinite, as for any eigenmode. With respect to the specific response function to be considered, one may think of calculating the density-density response function since in the low frequency limit the collective mode corresponds physically to sound waves [citer Randeria et travaux antŽrieurs]. Or else one may consider the anomalous response function since, on the BEC side for example, the collective mode coincides with the Bogoliubov excitations which are just the elementary excitations of the system out of the condensate. Actually both response functions lead naturally to the same answer for the collective mode, because density fluctuations are coupled to phase fluctuations of the order parameter since the particle current is proportional to the gradient of the phase. Naturally the response function is by itself quite interesting, not the least since its imaginary part is directly related to the dynamical structure factor, which has been studied very recently in the BEC-BCS crossover. We give also its explicit expression in the appendix for completeness.

In the following section we give the explicit expressions allowing to calculate numerically and study analytically the collective mode dispersion relation. Some details on its derivation are given in the appendix.
We then go on and study it in various situations along the BEC-BCS crossover.

\section{COLLECTIVE MODE DISPERSION RELATION}\label{disp}

From the kinetic equation or the diagrammatic method (see Appendix) we end up with the following dispersion relation, between the frequency $\omega $ of the collective mode and its wavevector $q$:
\begin{eqnarray}\label{eqdisprel}
{I}_{11}{I}_{22}=\omega ^{2}\,{I}^{2}_{12}
\end{eqnarray}
where the various integrals $I$ are defined by :
\begin{eqnarray}\label{I12}
{I}_{12}(\omega ,q)= \int_{0}^{\infty} \!\!{k}^{2}d{k} \int_{0}^{1}\!\!du\,\frac{1}{{E}_{+}{E}_{-}} \,\frac{{E}_{+}{\xi}_{-}+{\xi}_{+}{E}_{-}}{({E}_{+}+{E}_{-})^{2}-\omega ^{2}} 
\end{eqnarray}
\begin{eqnarray}\label{I11}
{I}_{11}(\omega ,q)= \int_{0}^{\infty} \!\!{k}^{2}d{k} \left[ \int_{0}^{1}\!\!du\,\frac{({E}_{+}+{E}_{-})}{{E}_{+}{E}_{-}} \,\frac{{E}_{+}{E}_{-}+{\xi}_{+}{\xi}_{-}+{\Delta}^{2}}{({E}_{+}+{E}_{-})^{2}-\omega ^{2}} - \frac{1}{{E}}\,\right ]
\end{eqnarray}
\begin{eqnarray}\label{I22}
{I}_{22}(\omega ,q)= \int_{0}^{\infty} \!\!{k}^{2}d{k} \left[ \int_{0}^{1}\!\!du\,\frac{({E}_{+}+{E}_{-})}{{E}_{+}{E}_{-}}\, \frac{{E}_{+}{E}_{-}+{\xi}_{+}{\xi}_{-}-{\Delta}^{2}}{({E}_{+}+{E}_{-})^{2}-\omega ^{2}} - \frac{1}{{E}}\,\right ]
\end{eqnarray}
and we have set ${E}_{\pm}=\sqrt{{\xi}_{\pm}^{2}+{\Delta}^{2}}$, ${E}=\sqrt{{\xi}^{2}+{\Delta}^{2}}$ with ${\xi}_{\pm}=({k}^2 \pm {k}qu + q^{2}/4)/(2m)-\mu $ and ${\xi}={k}^2/(2m)-\mu $. In the above integrals we have introduced the variable $u=\cos \theta$ where $\theta$ is the angle between ${\bf k}$ and ${\bf q}$.

Now it is quite interesting to note that one can also write:
\begin{eqnarray}\label{I11p}
{I}_{11}(\omega ,q)= \frac{1}{2} \int_{0}^{\infty} \!\!{k}^{2}d{k} \int_{0}^{1}\!\!du\,\frac{({E}_{+}+{E}_{-})}{{E}_{+}{E}_{-}} \,\frac{\omega^{2}-(q{k}u/m)^{2}}{({E}_{+}+{E}_{-})^{2}-\omega ^{2}} 
\end{eqnarray}
The identity between Eq.(\ref{I11}) and Eq.(\ref{I11p}) does not come by chance, but it is
just a consequence of mass conservation which is naturally satisfied by our starting equations. We could have used this mass conservation in the course of our derivation to  find directly Eq.(\ref{I11p}),
but it is here more efficient to just check explicitly the identity. 
Indeed this identity is equivalent to state that:
\begin{eqnarray}\label{}
 \sum_{k} \frac{(E_{+}+E_{-})}{E_{+}E_{-}} \,\frac{E_{+}E_{-}+\xi_{+}\xi_{-}+\Delta^{2}+(1/2)({\bf k}.{\bf q}/m)^{2}-\omega ^{2}/2}{(E_{+}+E_{-})^{2}-\omega ^{2}} - \frac{1}{E} =  \sum_{k} \frac{1}{2}(\frac{1}{E_+} +\frac{1}{E_-}) - \frac{1}{E} = 0
 \end{eqnarray}
 where the last equality is obtained by making appropriately the changes ${\bf k} \rightarrow {\bf k}\pm {\bf q}/4$ in the summation variable.
 
 Result (\ref{I11p}) for  ${I}_{11}(\omega ,q)$ makes it very easy to check that the dispersion relation (\ref{eqdisprel}) exhibits the proper phononic behaviour   $\omega=c_s q$ as $\omega$ and $q$ tend to $0$. 
  Indeed
 it this case we have ${I}_{12}(0,0)=(1/2)J_{\xi}$, ${I}_{22}(0,0)= - \Delta^{2} J_2$ and
 ${I}_{11}(\omega ,q)\simeq \omega^{2}J_2 /4-q^{2} J_4 /(12m^2)$ which leads immediately to
 the expression Eq.(\ref{eqcs}) for the sound velocity. 
 
\section{RESULTS}
\subsection{THRESHOLD FOR SINGLE FERMION EXCITATION SPECTRUM}
\label{thresh}

\begin{figure}\centering
\vbox to 90mm{ \epsfysize=90mm \epsfbox{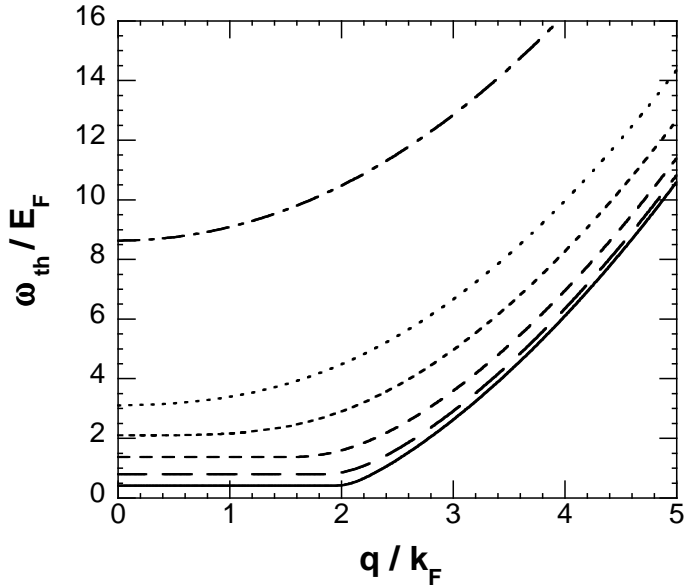}}
\caption{Threshold $\omega _{th}$ as a function of wavevector $q$ for, from bottom to top, $1/k_Fa=$ -1. , -0.5 , 0. (unitarity) , 0.553 ($\mu =0$) , 1. and 2.}
\label{excitnormtot}
\end{figure}

All the above integrals display singularities when $\omega={E}_{+}+{E}_{-}$. This equality corresponds physically to the possibility of breaking a Cooper pair into two fermionic excitations. The spectrum for this kind of excitation is continuous. The corresponding domain is bounded from below by the line $\omega_{th}={\mathrm min}\,({E}_{+}+{E}_{-})$ where the minimum is to be taken over all the possible values of ${\bf k}$. It is easy to see that the minimum is obtained for ${\bf k}.{\bf q}=0$ which gives ${E}_{+}={E}_{-}$. In the case $\mu >0$ and for $q \le 2 \sqrt{2m\mu }$, the minimum of ${E}_{+}+{E}_{-}$ is reached for ${k}^{2}/2m=\mu -q^{2}/8m$, so that $\omega_{th}=2 {\Delta}$. For $\mu >0$ and $q \ge 2 \sqrt{2m\mu }$, the minimum is reached for ${k}=0$, which leads to $\omega_{th}=2 \sqrt{(q^{2}/8m-\mu )^{2}+{\Delta}^{2}}$. On the other hand, for the case $\mu <0$, the minimum is always reached for ${k}=0$, leading to $\omega_{th}=2 \sqrt{(q^{2}/8m+|\mu | )^{2}+{\Delta}^{2}}$. To summarize we have:
\begin{eqnarray}\label{eqthres}
\omega_{th}=2 {\Delta} \hspace{20mm}{\mathrm for}\hspace{5mm} \mu  > 0 \hspace{5mm}{\mathrm and}\hspace{5mm} q \le 2\, \sqrt{2m\mu } \\
\nonumber
\omega_{th}=2 \sqrt{(q^{2}/8m-\mu )^{2}+{\Delta}^{2}} \hspace{28mm}{\mathrm otherwise}
\end{eqnarray}
A few examples of this threshold are given in Fig.\ref{excitnormtot}.
This result differs from the predictions of the non-interacting Fermi gas result, which has not only a lower boundary $\omega_{th-}=0$ for $q \le 2 k_F$, and $\omega_{th-}=-q k_F/m+q^{2}/2m$ for $q \ge 2 k_F$, but also an upper boundary $\omega_{th+}=q k_F/m+q^{2}/2m$. It is worth noticing that this behaviour is not recovered in the  ${\Delta} \rightarrow 0$ limit reflecting the fact that, with respect to the excitation spectrum domain, the normal state limit is singular. However one recovers of course in this limit the correct behaviour of the dynamic structure factor, because outside the standard normal state domain the structure factor will go to zero, due to the effect of coherence factors.

In the BEC limit, one has ${\Delta}/\mu  \rightarrow 0_{-}$ which gives $\omega_{th}=2 |\mu | + q^{2}/4m$. The physical meaning is clear. This is just the energy necessary to break a molecule with binding energy $\epsilon _b =  2 |\mu | = 1/ma^2$ into two fermions with total momentum ${\bf q}$, leading to an additional kinetic energy $(1/2) q^{2}/(2m)$, just as for a molecule of mass $2m$.

\subsection{WEAK COUPLING BCS REGIME}
\label{wc}

\begin{figure}\centering
\vbox to 90mm{ \epsfysize=90mm \epsfbox{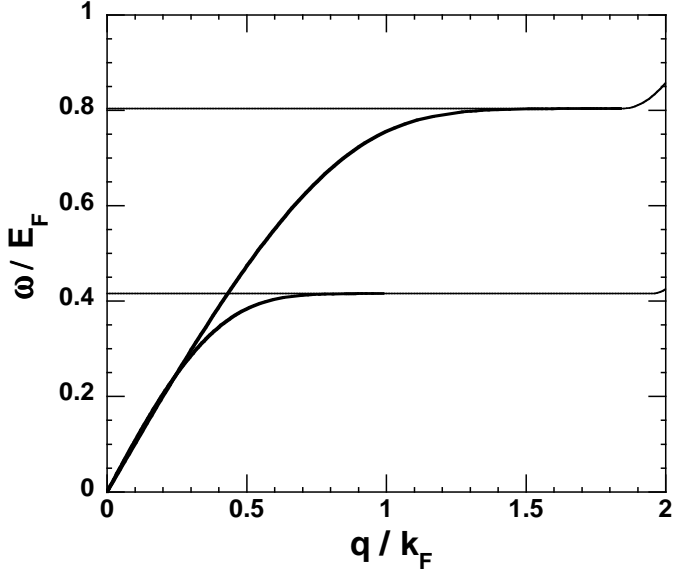}}
\caption{Dispersion relation $\omega /E_F$ of the collective mode as a function of $q/k_F$ for $1/k_Fa= -1. $ (lower thick line) and $-0.5$ (upper thick line). The location (see section \ref{thresh}) of the threshold for pair-breaking is given in each case by the thiner line.}
\label{dispersexcit}
\end{figure}

Let us consider first the weak coupling regime on the BCS side, where $a \rightarrow 0_-$ and $\Delta \rightarrow 0$. The numerical result for $1/k_Fa = -1.$ and  $-0.5$ are displayed on Fig.\ref{dispersexcit}. Although in these cases $a$ is not so small, they still display a behaviour which is typical of this limit and which would just be more strongly marked at smaller $a$. For increasing $q$ the dispersion relation has first a fairly extended linear part, the slope being naturally the sound velocity. Then it bends to become nearly horizontal. This last behaviour is quite reminiscent of an anticrossing in a two level system. Here the collective mode plays the role of one of the levels. The other one is not really a level since it is rather the pair-breaking continuum. However the well-known divergence  of the BCS density of states at the  $\omega = 2 \Delta $ threshold  makes the threshold level quite similar to a single level at energy $\omega = 2 \Delta $. It is quite interesting to note that this behaviour of the dispersion relation of the collective mode is a unique feature of the ultracold Fermi gas system. Indeed in superconductors the collective mode frequency is pushed up to the plasma frequency because the paired electrons are charged particles. On the other hand, in superfluid $^3He$ which is a neutral BCS superfluid, the strong hard core repulsion between atoms produce a much faster sound velocity and the mode  actually merges in the pair-breaking continuum. Hence the anticrossing feature is found only in Fermi gases.

Looking at Fig.\ref{dispersexcit} one would wrongly conclude that the collective mode touches the pair-breaking continuum and merges with it for $q < 2 \sqrt{2m\mu }$.  This is not the case, as the above anticrossing argument makes it clear and can be analytically shown (see the Appendix). 
As we will see in the next section \ref{contact}, the collective mode does indeed merge in the continuum in the weak coupling regime, but this always occurs for $q > 2 \sqrt{2m\mu }$,  the merging point approaching the $q = 2 \sqrt{2m\mu }$ point as the coupling becomes weaker and weaker.

\subsection{CONTACT OF THE COLLECTIVE MODE WITH THE CONTINUUM}
\label{contact}

\begin{figure}\centering
\hspace*{20mm}
\vbox to 90mm{\epsfysize=90mm \epsfbox{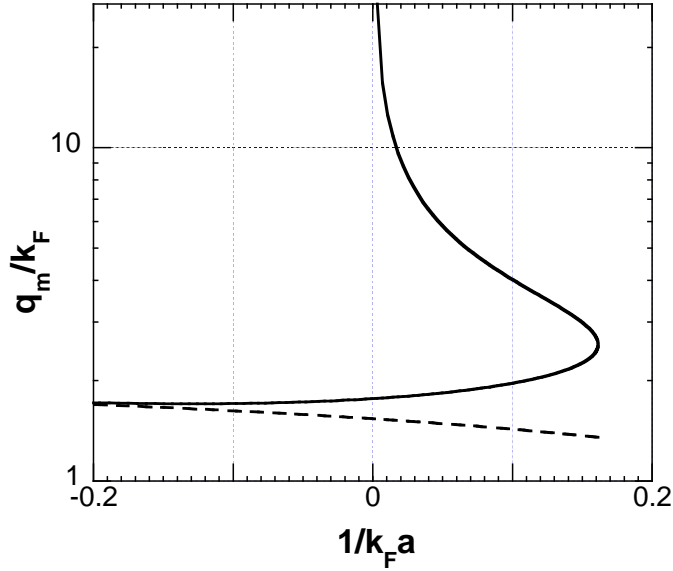}}
\caption{Wavevector $q_m$, in units of $k_F$, at which the collective mode dispersion relation meets the excitation continuum, as a function of $1/k_Fa$ (note that the $q_m$ scale is logarithmic). The dashed line corresponds to $q =2 \sqrt{2m\mu }$. It goes to $q/ k_F=2$ when $a \rightarrow - \infty$}
\label{figqm}
\end{figure}

In order to clarify the position of the collective mode dispersion relation with respect to the threshold of
the continuous excitation spectrum, it is useful to find precisely the wavevector $q_m$ (and corresponding frequency $\omega _m$) at which they meet. Since the meeting never occurs for $\mu  <0$
one can set $\omega=2 \sqrt{(q^{2}/8m-\mu )^{2}+{\Delta}^{2}}$ with $\mu >0$ in Eq.(\ref{eqdisprel}) and solve the corresponding equation numerically. The result is displayed on Fig. \ref{figqm}.

First of all, it is clear from these results that, as soon as $a < 0$, the mode meets the continuum slightly beyond the point $\omega = 2 \Delta $, and $q=2 \sqrt{2m\mu }$, where the lower boundary of the continuum begins to depart from the simple value $\omega_{th} = 2 \Delta $. The corresponding wavevector is indicated as a dashed line on Fig. \ref{figqm}. 
When one goes towards large and negative values of $a$, the meeting point ($q_m,\omega_m$) gets extremely close to $(2 \sqrt{2m\mu }, 2 \Delta)$, while never reaching it exactly.
The meeting point begins to depart appreciably from this location in the $q-\omega$ plane only when one gets close to unitarity. However at this stage a rather surprising feature appears. On the other side of the resonance, for $a>0$, a second branch of the dispersion relation for the collective mode takes place  at  very large values of $q$, extending up to $q\rightarrow \infty$. The appearance of this large $q$ branch at unitarity can be confirmed analytically. This branch meets the continuum at some lower value $q_m$. This means that the above equation has two solutions instead of one, as it was the case for $a < 0$. This second solution corresponds to the upper branch of the curve seen in Fig. \ref{figqm}. When one goes away from unitarity on the $a > 0$ side, the two meeting points move rapidly towards each other. They are found numerically to merge into a double solution $q/k_F=2.57$ for $1/k_Fa=0.161$.

Beyond this value of $1/k_Fa$, the above equation has no longer solutions, which means that the collective mode never meets the continuum. Hence, qualitatively, one has already the situation found far on the BEC side. It is remarkable that this evolution occurs very rapidly, as a function of the scattering length, since this happens basically between $1/k_Fa=0$ and $1/k_Fa=0.16$.
It is also quite noticeable that this change occurs while the chemical potential is still positive, a regime where one would not be tempted to speak of molecules since the molecular formation would seemingly imply a negative chemical potential. Nevertheless in this regime we find a fully characterized collective mode, similar to the elementary excitations of the molecular condensate.
It is quite tempting to interpret the remarkable fact that the collective branch meets the continuum twice for large and positive values of the scattering length $a$ in the following way. It has been shown \cite{rcmol,clk} that, in the normal state, the presence of the Fermi sea of atoms makes it more difficult for molecules to form, because some states are blocked due to Pauli principle by the presence of other atoms. This is mostly felt for molecules with low total momentum, which do not form at unitarity, but rather only at some positive value of $a$ depending on temperature. On the other hand molecules with very high momentum are insensitive to the presence of the Fermi sea and form right at unitarity. Now, for a standard Bose-Einstein condensate, the physical nature of the collective mode at very high $q$ is just the  excitation of a single boson of momentum $q$ out of the condensate. Similarly for the Bose condensate of fermionic molecules occuring in our full BEC regime, the collective mode at very high $q$ is the excitation out of the condensate of a molecule of momentum $q$. The natural extension of the above result, regarding molecular formation in the normal state, is to say that, just for the same physical reasons, these very high $q$ excited molecules form also unhindered in the superfluid state at $T=0$. Hence for very high $q$ the collective mode exists as long as these molecules exist, that is up to unitarity when we increase $a$. In other words for positive values of $a$ and large $q$ one always expect to find a discretized branch due to the fact that in this case the two-body problem always admits the existence of a molecular state and many body effects are negligible. Naturally the difference between this excitation and molecular breaking, which corresponds to the lower boundary of the continuum, decreases when $a$ increases and at unitarity, where the molecular binding energy becomes zero, the collective mode merges into the continuum. On the other hand when we decrease $q$, we expect that the formation of excited molecules will become more difficult due to Pauli exclusion by the condensate, because the Pauli principle becomes more and more effective, reducing effectively in this way the binding energy. This explains why at some stage the collective mode merges into the continuum when $q$ is decreased. Let us stress that, if this physical interpretation is correct, this implies immediately that the exact theory of the BEC-BCS crossover should also display this behaviour because our arguments do not make use specifically of the BCS theory. This means that we exclude a scenario where the full curve in Fig. \ref{figqm} goes to infinity as one approaches the resonance from the negative $a$ side, which would result in this curve being entirely located on the $a<0$ side. Clearly it would be worthwhile to investigate experimentally this question.

\subsection{BEC LIMIT AND HIGH FREQUENCY LIMIT}
\label{bec}

We expect naturally to find the Bogoliubov mode when we go to the BEC limit, which corresponds to $\mu  < 0$ and $a \rightarrow 0_+$. Physically we have in this case a dilute gas of very tight molecules, which should behave as elementary bosons. In this limit the well-known dispersion relation of the Bogoliubov mode is given by:
\begin{eqnarray}\label{eqbog}
\omega ^2 = c_{s}^{2} q^2 + \left( \frac{q^2}{2M_m} \right)^2
\end{eqnarray}
where, as we have seen, the sound velocity $c_s$ is given by $c^{2}_s=g_m n_m/M_m=2\pi na / m^2$ and $M_m=2m$ is the molecular mass. Since we are in the $a \rightarrow 0_+$ limit, the first term will be negligible compared to the second one. More precisely this means that we will consider wavevectors $q \gg (na)^{1/2}$, i.e. much larger than the inverse of the healing length of the molecular Bose gas. In this case we will have merely $\omega \simeq q^2/4m$ and we are in the regime where the mode corresponds physically to kick a molecule out of the condensate.

In this BEC limit the gas parameter $na^3$ becomes very small and ${\Delta }/|\mu | \sim (na^3)^{1/2} \rightarrow 0$. On the other hand the above condition $q \gg (na)^{1/2}$ is equivalent to $q^2/2m \gg \Delta$. Hence, in this limit $\Delta $ becomes negligible compared to all the other energies in the problem, which means that we have to take it formally going to zero.
In this situation, we have merely to take ${E}_{±\pm}={\xi}_{±\pm}$ and the integrals in Eq.(\ref{eqdisprel}) can all be easily performed analytically. One finds:
\begin{eqnarray}\label{eqgdi12bec}
{I}_{12} = (2m)^{3/2} \frac{\pi }{8 \omega } \left[ \sqrt{4|\mu |  + q^2/2m + 2 \omega} - \sqrt{4|\mu | + q^2/2m - 2 \omega}\;\right]
\end{eqnarray}
\begin{eqnarray}\label{eqgdi11bec}
{I}_{11}={I}_{22} =(2m)^{3/2} \frac{\pi }{8 } \left[ \,4|\mu | ^{1/2}- \sqrt{4|\mu |  + q^2/2m + 2 \omega} - \sqrt{4|\mu | + q^2/2m - 2 \omega}\;\right]
\end{eqnarray}
Then Eq.(\ref{eqdisprel}) becomes ${I}_{11}+\omega {I}_{12}=0$ (since $ {I}_{11} < 0$) and one finds immediately that the solution is indeed $\omega = q^2/4m$.

We note also that, if we investigate the regime where $q$ and $\omega $ go to infinity whatever the value of  $na^3$ (provided $\mu < 0$), we are lead to perform exactly the same calculation. Indeed in this limit, $\omega $ and $q^2/2m$ are large compared to $\Delta $, so we can take again $\Delta \rightarrow 0$. Therefore the result for the dispersion relation is also $\omega = q^2/4m$. This is again in agreement with the physical picture holding for the mode in this limit, namely kicking a molecule out of the condensate.

\subsection{THE COLLECTIVE MODE AT UNITARITY AND ON THE BEC SIDE OF THE CROSSOVER}
\label{unitbec}
 
\begin{figure}\centering
\vbox to 90mm{\epsfysize=90mm \epsfbox{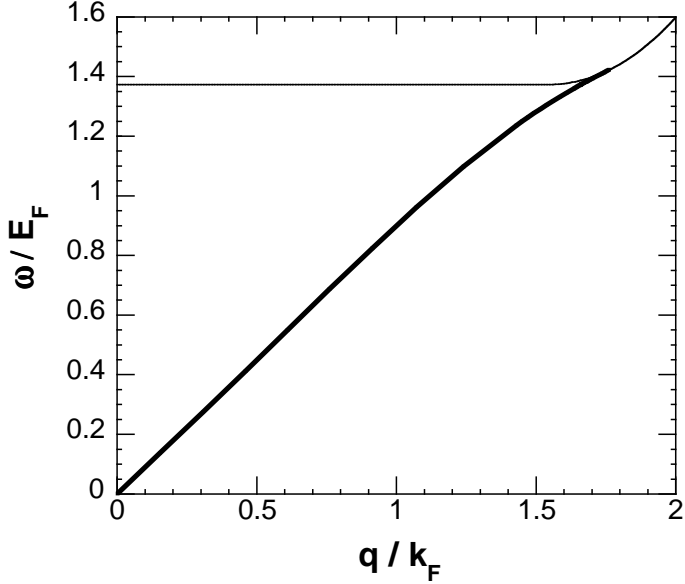}}
\caption{Dispersion relation $\omega /E_F$ of the collective mode as a function of $q/k_F$ at unitarity (thick line). The location (see section \ref{thresh}) of the threshold for pair-breaking is given by the thiner line. The collective mode merges into the continuum for $q/k_F=1.76$, as it can be seen from Fig.\ref{figqm}}
\label{figunit}
\end{figure}

Now we will proceed to give results for the collective mode dispersion relation on the BEC side
of the crossover. We first exhibit the result right at unitarity (see Fig.\ref{figunit}). It is somewhat analogous to what has already been found on the weak coupling side in section \ref{wc}, with the dispersion relation merging into the continuum for $q/k_F=1.76$.

A noticeable feature of Fig.\ref{figunit} is that the dispersion relation is surprisingly almost linear up to the merging with the continuum. This result predicted by the BCS model can be relevant for understanding the experimental data on the collective oscillations in trapped Fermi gases. In fact in actual experiments carried out in elongated traps at unitarity the superfluid gap $\Delta$ is never much larger than the radial oscillator frequency $\omega_\perp$ which, within hydrodymamics, fixes the collective frequency of the radial breathing mode according to the law \cite{strepl} $\omega_{rad}=\sqrt{10/3}\,\omega_\perp$. This means that in these experiments one is exploring a region of relatively high wavevectors. The quasi-linearity up to high wavevectors allows one to understand why hydrodynamics (valid in principle at low wavevectors) is in such a good agreement with experiments \cite{grimnew,note}. A careful study of the dependence of the collective frequency on the ratio $\Delta/\omega_{rad}$ might actually provide new insight on the behaviour of the excitation spectrum and in particular check if the extended linearity of the dispersion law is a peculiarity of the BCS model or is a more general feature.

\begin{figure}\centering
\hspace*{20mm}
\vbox to 90mm{\epsfysize=90mm \epsfbox{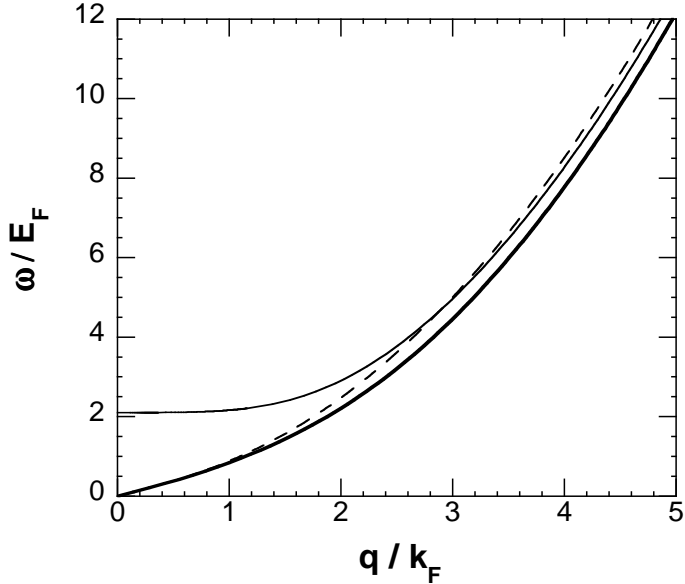}}
\caption{Dispersion relation $\omega /E_F$ of the collective mode  (thick line) as a function of $q/k_F$ for $\mu =0$, corresponding to $1/k_Fa  = 0.553$. The location (see section \ref{thresh}) of the threshold for pair-breaking is given by the thiner line. The dashed line indicates the result obtained from the Bogoliubov formula for the collective mode for this value of $\mu $.}
\label{figmu0}
\end{figure}

On Fig. \ref{figmu0} we display the results when $ \mu =0$, corresponding to $1/k_Fa  = 0.553$ as we have seen in section \ref{mf}. Since we are beyond the value $1/k_Fa  = 0.161$, the dispersion relation is now fully below the continuum as discussed in section \ref{contact}. However for most of the range it is quite close to lower boundary of this continuum, and the situation is not drastically different from the one obtained at unitarity. This is even more so if we take into account that the spectral weight associated with the collective mode gets smaller when it goes near the continuum. On the other hand the shape of the dispersion relation is already quite similar to the familiar one given by the Bogoliubov result, already discussed in section \ref{bec}. We have made this clear by plotting the result of the Bogoliubov dispersion relation in the same figure, with the sound velocity $c_s$ taken naturally for $\mu =0$. Although we are certainly not in the deep BEC regime where this formula is expected to be valid, we see that it is quite similar to the actual result, although it goes unappropriately in the continuum.
The situation becomes quite different when we move slightly deeper in the BEC regime as it can be seen in Fig.\ref{figmoins1}, where we have plotted the results for $1/k_Fa  = 1.$. We see that the collective mode has moved much further from the continuum. At the same time the actual result is quite closely approximated by the Bogoliubov formula. Hence it is not surprising that, when we go to $1/k_Fa  = 2.$, the Bogoliubov result becomes almost undistinguishable from our numerical result, as it can be seen in Fig.\ref{figmoins2}, which shows that we are already effectively quite deep in the BEC regime. This is coherent with the fact that the absolute value $|\mu |$ of the chemical potential is already almost 4 times the Fermi energy $E_F$, as it can be seen from Ref. \cite{sdm,gss}. Noting that for the dilute limit, this chemical potential is half the free molecular binding energy $|\mu |= 1/2ma^2$, leading to $|\mu |/E_F = 1/(k_Fa)^2 = 4$, one sees that one is indeed already quite near this situation.
This excellent agreement between the BCS result and the Bogoliubov formula makes it worthwhile to investigate the collective mode dispersion relation in the dilute limit to next order in powers of $\Delta /|\mu | $, that is to first order in $(\Delta /|\mu |)^2$. This means going one step further than what we have done in the preceding section \ref{bec}. This turns out to be not so complicated analytically. Details of the derivation of this expansion are given in the appendix.
One obtains:
\begin{eqnarray}\label{eqdispdev1}
\omega^2 =  (qa)^2\;\frac{{ \Delta }^2}{4}\,\frac{16-(qa)^2}{16+(qa)^2}+\left( \frac{q^2}{4m} \right)^2
\end{eqnarray}
up to first order in $(\Delta /|\mu |)^2 \sim na^3$, where we have used $|\mu |=1/(2ma^2)$ to lowest order. This has to be compared with the result from the Bogoliubov formula Eq.(\ref{eqbog}).
Taking the sound velocity $c_{s}^{2}={ \Delta}^2/(8m|\mu |)$ in the BEC limit from section \ref{mf}, we see that, in the limit $qa \ll 1$, this is in perfect agreement with Eq.(\ref{eqdispdev1}) and in particular we recover the well known result
\begin{eqnarray}\label{shiftmu1}
 \omega = \frac{q^2}{4m} + \frac{4\pi na}{m}
 \end{eqnarray}
which expresses the {\it positive} shift of the dispersion with respect to the free value $q^2/4m$ in terms of the scattering length.  
On the other hand, in the limit $qa \gg 1$, the dispersion relation becomes instead:
\begin{eqnarray}\label{shiftmu2}
\omega = \frac{q^2}{4m} - \frac{4\pi na}{m}
\end{eqnarray}
showing that in this case the shift has a {\it negative} sign. These results are in qualitative agreement with our numerical results seen in Fig. \ref{figmoins1} and \ref{figmoins2}. Indeed these figures show that the agreement with the Boboliubov result (which is indeed above the numerical result) is not so good for large $qa$, whereas it is remarkably good for lower $qa$ (note that in the linear range $qa \ll { \Delta}/|\mu |$ we have automatically agreement since the sound velocity is the same).
This result is physically reasonable. Indeed, for $q \ll 1/a$, one is testing the system at lengths which are larger than the molecular size $a$. Hence it is natural that one gets the behaviour of pure bosons and an agreement with Bogoliubov theory.
On the other hand for $q \gg 1/a$ one is testing lengths which are smaller than the molecular size, and the fact that we deal with composite bosons appear at this scale. It is therefore natural to find a disagreement with Bogoliubov theory which does not include this physics.

 \subsection{Critical velocity}

We have now basically all the informations required to calculate the critical velocity of the superfluid $v_c$ in the BEC-BCS crossover. Indeed, according to Landau's criterion, it is given by $v_c = {\rm Min}(\omega (q)/q)$ where $\omega (q)$ is the energy of an elementary excitation. For this BCS superfluid we have two kind of excitations. First we have bosonic excitations corresponding to the collective mode we are studying. If the dispersion relation has an upward curvature, the minimum of $\omega (q)/q)$ is obtained for $q\rightarrow 0$, which gives for $v_c$ the sound velocity $v_s$ calculated in section \ref{vs} if we retain only this kind of excitations. However we have also to take into account fermionic single particle excitations $\omega (k)=\sqrt{\xi_{k}^{2}+\Delta ^{2}}$, and we have to find again the minimum of $\omega (k)/k$. However we have just seen that the lower boundary for the pair-breaking continuum corresponds to the creation of two single particle excitations with same energies and wavevectors, since we have found ${E}_{+}={E}_{-}$ in this case. Hence it is easily seen that, since the wavevector corresponding to the minimum will clearly be greater than $\sqrt{2m|\mu|}$, ${\rm Min}(\omega (k)/k)$ for single particle excitations is identical to ${\rm Min}(\omega_{th} (q)/q)$ for the pair-breaking continuum found in Fig.\ref{excitnormtot} (the correspondence is just obtained by multiplying energies and wavevectors by a factor 2). This minimum is easily found analytically to be given by $([\sqrt{\Delta ^{2}+\mu ^{2}}-\mu]/m)^{1/2}$, obtained for the single particle wavevector $ [4m^2(\Delta ^{2}+\mu ^{2})]^{1/4}$. Hence, taking into account both kind of excitations, the critical velocity is given by:
\begin{eqnarray}\label{vcrit}
v_c={\rm Min}\left(v_s,([\sqrt{\Delta ^{2}+\mu ^{2}}-\mu]/m)^{1/2}\right)
\end{eqnarray}
which is plotted in Fig.\ref{figvc}. The result displays a kink at its maximum, occuring when one switches from bosonic to fermionic excitations. It occurs very near unitarity, on the BEC side. It is worth noticing that, in terms of critical velocity, the strength of the superfluid is at its highest around unitarity and not on the BEC side, as one might perhaps think at first.

\begin{figure}\centering
\vbox to 90mm{\epsfysize=90mm \epsfbox{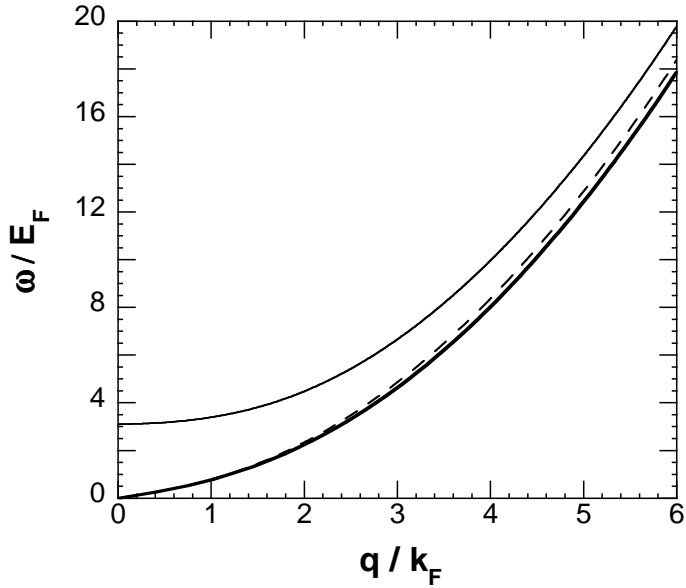}}
\caption{Dispersion relation $\omega /E_F$ of the collective mode as a function of $q/k_F$ for $1/k_Fa  = 1.$ (thick line), for which $ \Delta /|\mu | = 1.66$. The threshold for pair-breaking is given by the thiner line. The dashed line indicates the result obtained from the Bogoliubov formula.}
\label{figmoins1}
\end{figure}

\begin{figure}\centering
\vbox to 90mm{\epsfysize=90mm \epsfbox{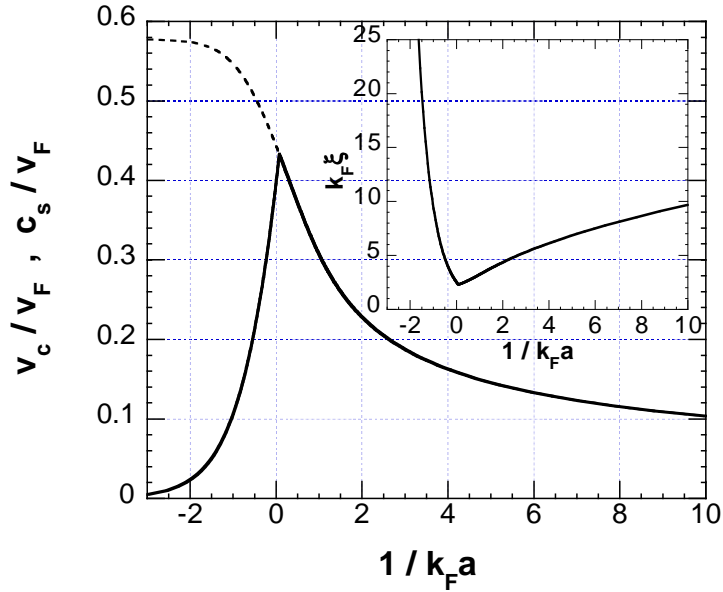}}
\caption{Critical velocity $v_c$ (full line) and sound velocity $c_s$ (dashed line) as a function of $1/k_Fa$. The insert gives $k_F\xi=v_F/v_c$.}
\label{figvc}
\end{figure}

\begin{figure}\centering
\vbox to 90mm{\epsfysize=90mm \epsfbox{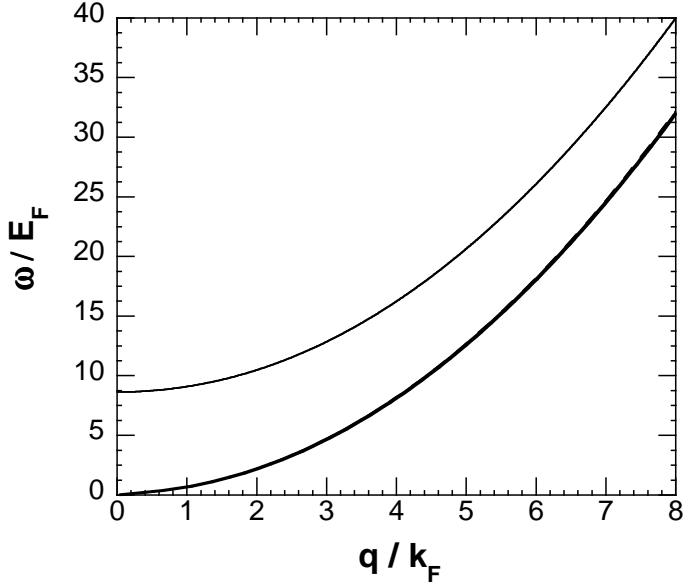}}
\caption{Dispersion relation $\omega /E_F$ of the collective mode as a function of $q/k_F$ for $1/k_Fa  = 2.$ (thick line), for which $ \Delta /|\mu | = 0.474$.The threshold for pair-breaking is given by the thiner line. The dashed line indicates the result obtained from the Bogoliubov formula.}
\label{figmoins2}
\end{figure}

In the above determination of the critical velocity, we have considered that the collective mode dispersion relation is bending upward. However, while this is clearly so when the chemical potential is negative (see Fig.\ref{figmoins2}, \ref{figmoins1} and \ref{figmu0}), this is no longer true at unitarity (see Fig.\ref{figunit}) and on the BCS side (see Fig. \ref{dispersexcit}). Nevertheless these last situations are irrelevant for our purpose since in these cases, the critical velocity is due to single particle excitations. But one may still wonder if, near unitarity, one does meet the case where some part at least of the collective mode dispersion relation would display downward bending, which could result in ${\rm Min}(\omega (q)/q)$ obtained not for $q\rightarrow 0$, but for some finite value of $q$, as it occurs for example in superfluid $^4$He for the roton minimum. To clear up this point, we have investigated in detail the collective mode dispersion relation in the vicinity of the maximum of $v_c$. Indeed the switch between the behaviour found in Fig.\ref{figunit} and in Fig.\ref{figmu0} occurs by an upward bending appearing in the low $q$ region, while the higher $q$ part has still a downward bending. However for this higher part we find always that $\omega (q)/q > c_s$ so it does not play any role in the determination of the critical velocity, and accordingly there is no modification to bring to Fig.\ref{figvc}, and in particular the kink is not removed. It must also stressed that, for all this region near unitarity, these bendings (upward or downward) are extremely small, and the most striking feature is that the dispersion relation is remarkably essentially a straight line between $q=0$ and its lowest merging with the continuum. This is particularly so in the immediate vicinity of the maximum of $v_c$ where the difference with a straight line is quite minute and could not be seen on a figure (which accordingly spare us from displaying it). Nevertheless the numerical calculation shows that this is never exactly a straight line. It would clearly be very interesting to know if this peculiar and remarkable behaviour is just a coincidental specificity of the BCS model, or if it has a more general validity.
An important physical quantity directly related to $v_c$ is the healing length, which we define merely as:
\begin{eqnarray}\label{}
\xi = \frac{1}{mv_c}
\end{eqnarray}
It has been studied in details by Pistolesi and Strinati \cite{pistr}. We have plotted it in the insert of Fig.\ref{figvc}. It coincides, apart from a trivial numerical factor, with the usual definition of healing length in the molecular BEC limit and with the size of Cooper pairs, i.e. the coherence length, in the opposite BCS limit. The healing length, which takes its smallest value very near unitarity, fixes the core size of quantized vortices.
 
\section{CONCLUSION}

In this paper we have made a detailed study of the collective mode across the whole BEC-BCS crossover in fermionic gases at zero temperature. This has been done on the basis of the dynamical BCS model, which we have formulated equivalently in terms of diagrammatic theory or with collisionless kinetic equation. Naturally we have first recovered the results of the linear regime, where the mode is physically identical to sound propagation. We have cast the result for sound velocity in a simple form and payed particular attention to the vicinity of the point where the chemical potential is zero. Then we have turned to the non linear part of the dispersion relation. Our investigation has revealed interesting qualitative behaviours which would deserve further exploration with better theoretical tools, and would also be worthwhile to investigate experimentally. Quite generally there is an interplay between the collective mode and the continuum of single fermionic excitations. Roughly speaking the behaviour is analogous to the anticrossing phenomenon well-known for two level systems. This behaviour is most clearly displayed on the BCS side, which is quite interesting in itself since it provides the first experimental example of a pure Anderson-Bogoliubov mode. Nevertheless this anticrossing behaviour does not prevent the collective mode to merge at some stage into the single fermionic excitations continuum. This occurs mostly on the BCS side of the crossover, but quite interestingly it exists also slightly beyond unitarity on the BEC side, where we have made a careful study. This feature might be related to the physics of molecular formation in the crossover. Another remarkable feature, which is coherent with experiments on collective excitations in trapped gases, is the very linear behaviour of the dispersion relation in the vicinity of unitarity almost up to merging with the continuum. Finally on the BEC side the mode is naturally quite analogous to the Bogoliubov mode and no merging occurs anymore. However, even deep in this BEC regime, there is at high wavevectors a difference between this mode and the result from the BCS model, which can most likely be ascribed to the composite nature of the molecular bosons when they are tested at length scales comparable to their size. Finally our results on the collective mode and on the single particle excitations allows us to obtain the Landau critical velocity, which display a sharp maximum very near unitarity.

\section{ACKNOWLEDGEMENTS}
M. Yu. K. is grateful to the Landau-ENS collaboration program for the support during the final stage of this work. He also acknowledges the financial support from the Russian Foundation for Basic Research (RFBR, grant N$^{\circ}$ 06-02-16449). Laboratoire de Physique Statistique is "associ\'e au Centre National
de la Recherche Scientifique et aux Universit\'es Pierre et Marie Curie-Paris6 et Paris 7"

\section{APPENDICES}
\subsection{Vicinity of $\mu =0$}
We give the results of the first order expansion in the vinicity of $\mu =0$. All the specific numerical integrals coming in this calculation are elliptic integrals \cite{grry}, which can all be expressed in this case in terms of Euler's gamma function $\Gamma (1/4)\simeq 3.62561 \equiv {\bar \Gamma}$. One finds:
\begin{eqnarray}\label{eqmu0}
\frac{1}{a\sqrt{2m\Delta }}= A - D \;\frac{\mu }{\Delta }   \hspace{12mm}
\left(\frac{E_F}{\Delta }\right)^{3/2}= B + C\; \frac{\mu }{\Delta }
\end{eqnarray}
where $A=4 \pi ^{1/2}/{\bar \Gamma}^2 \simeq 0.5393 $, $B=1/(2A) \simeq 0.9270 $, $C= 3 \pi A/8 \simeq 0.6354 $ and $D= 1/(\pi A) \simeq 0.5902 $. In particular one finds that, for $\mu =0$, $\Delta=\Delta_0$ with $\Delta_0 /E_F=4 \pi ^{1/3}/{\bar \Gamma}^{4/3} \simeq 1.0518$, with a corresponding value:
\begin{eqnarray}\label{}
\left(\frac{1}{k_Fa}\right)_{0} = \frac{8 \pi ^{2/3}}{{\bar \Gamma}^{8/3}} = \frac{1}{2}\,\left(\frac{\Delta_0}{E_F}\right)^2 \simeq 0.553
\end{eqnarray}

In the vicinity of $a_0$, we have from Eq.(\ref{eqmu0}):
\begin{eqnarray}\label{}
\Delta - \Delta _0 = - \frac{\pi A^2}{2} \mu  \simeq - 0.4569 \,\mu   \hspace{12mm}
\frac{1}{k_Fa}-\left(\frac{1}{k_Fa}\right)_{0}= - (1+\pi ^2 A^4/4)\,\frac{B^{1/3}}{\pi A} \,\frac{\mu }{E_F} \simeq - 0.6956  \,\frac{\mu }{E_F}
\end{eqnarray}

\subsection{COLLECTIVE MODE AND  RESPONSE FUNCTION}\label{resp}

In order to obtain the response function various approaches can be used, which are actually all basically equivalent presentations of the standard self-consistent BCS theory. We will consider two of them. One is the Green's function, or diagrammatic, approach to the BCS theory. The other one is kinetic theory. In the following subsection we will use the Green's function approach to calculate the response, and more precisely the condition under which the vertex diverges. Then in the next subsection we will use the kinetic equation method to obtain the condition of divergence of the density response. Naturally each one of this method can be applied to the calculation of the full response, which we will give below. In both cases we will be rather sketchy since these methods are well-known in the literature, and we give only the following indications for completeness.

\subsubsection{DIAGRAMMATIC APPROACH}

We will obtain the vertex
$\Gamma$ in the superfluid state
by writing a Bethe-Salpeter equation \cite{agd}. This vertex is a generalized 
T-matrix. In the present case it depends only on the total momentum
${\bf q} $ and frequency $\omega $, as can be seen in the explicit setup
of the Bethe-Salpeter equation, done below. More specifically we look for a pole of
this vertex. We introduce the normal and anomalous Green's functions, $G$ and
$F$ respectively, 
in the superfluid phase. It is more convenient to use the
Euclidean form (after Wick transformation $\omega \rightarrow i\omega$) 
for which we have $G (i\omega,
{\bf k}) =-(i\omega + \xi_k)/(\omega^2 + E_{k}^2)$ and
$F(i\omega, {\bf k}) = -\Delta / (\omega^2 +
E_{k}^2)$.

\begin{figure}\centering
\vbox to 40mm{\epsfysize=40mm \epsfbox{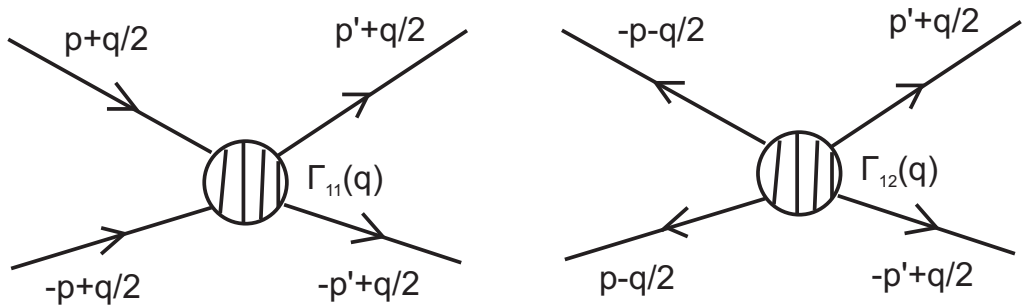}}
\caption{Two vertices $\Gamma_{11}(q)$ and $\Gamma_{12}(q)$ in
superconductive state.  $p=(\bar{p},\omega)$ and $q= (\Omega,
\bar{q})$ are 4-momenta} \label{fig1bis}
\end{figure}

\begin{figure}\centering
\vbox to 60mm{\epsfysize=60mm \epsfbox{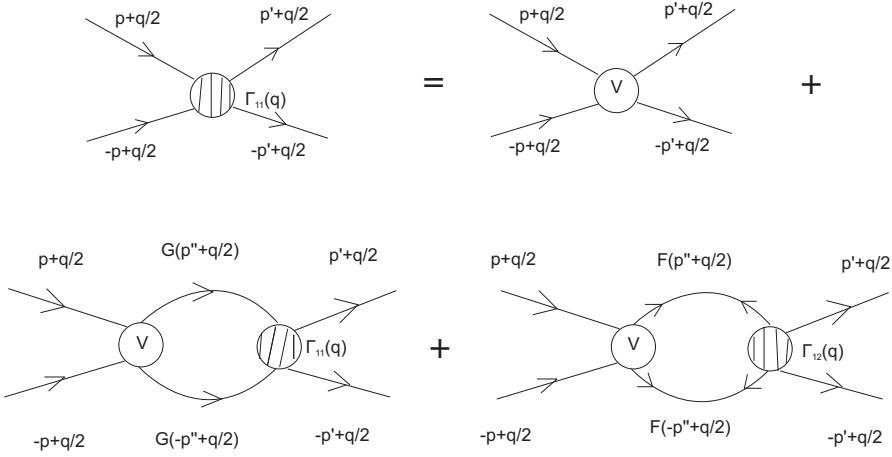}}
\caption{The Bethe-Salpeter equation for the vertex
$\Gamma_{11}$} \label{fig2bis}
\end{figure}

In order to obtain a Bethe-Salpeter equation, we have to
introduce two vertices $\Gamma_{11}$ and $\Gamma_{12}$ (see Fig.
\ref{fig1bis}). The first one $\Gamma_{11}$ corresponds to the scattering of two atoms with
opposite spins (with two in-going lines and two out-going ones), while the
second one $\Gamma_{12}$ has four out-going lines. This last one is non zero
only in the superfluid state, while $\Gamma_{11}$ exists also in the normal state.
We obtain the Bethe-Salpeter equation by considering first the simplest process
corresponding to an elementary interaction, depicted as the first term in the right-hand
side of Fig. \ref{fig2bis}. Then we consider more complicated processes obtained
by repeating this process, producing in this way ladder diagrams. In the intermediate
steps we can introduce normal Green's functions, as in the second term in the right-hand
side of Fig. \ref{fig2bis} (they are depicted by lines with single arrows). But we can
also introduce anomalous Green's functions, as in the third term in the right-hand
side of Fig. \ref{fig2bis} (they are depicted by lines with double arrows having opposite
directions).  On the other hand we can not have an  intermediate step made of one normal propagator and an anomalous one because this would imply interaction between atoms having the same spin, which is forbidden for s-wave scattering by Pauli principle. Summing up all the processes after the first one
lead to the equation represented diagrammatically in Fig. \ref{fig2bis}.
In algebraic form it reads
\begin{eqnarray}\label{eqbethsalp}
\Gamma_{11}(q)=V-V\,\Gamma_{11}(q)\int \!\frac{d^4p''}{(2\pi)^4} \,
G(p''+q/2)G(-p''+q/2)-V\,\Gamma_{12}(q)\int \!\frac{d^4p''}{(2\pi)^4} \,
F(p''+q/2)F(-p''+q/2) \label{eq1}
\end{eqnarray}
where $q$ is now a four-vector representing ($\omega , {\bf q} $).
If we want to take "renormalization" into account, as we have done in section \ref{mf}
when going from Eq.(\ref{eqga}) to Eq.(\ref{eqgap}), it is easy to show that we should
replace $G G$ in Eq.(\ref{eq1}) by $GG-G_0G_0$, where
$G_0(\omega,{\bf k})=[i\omega-\varepsilon_k]^{-1}$ is the vacuum
Green's function. Simultaneously we have to replace the bare interaction $V$
by the coupling constant $g=4\pi a/m$.
From Eq.(\ref{eqbethsalp}) it is natural to introduce now the elementary response functions $\chi_{ij}$
corresponding to the various bubbles appearing in Fig. \ref{fig2bis}. Specifically we define:
\begin{eqnarray}
\left\{
\begin{array}{lcl}
-\chi_{11}(q)=\frac{1}{g}+\int \! \frac{d^4p}{(2\pi)^4} [G(p+q/2)G(-p+q/2)-G_0(p)G_0(-p)] \\
\hspace{10mm}\\
\hspace{2mm}\chi_{12}(q)=\int \!\frac{d^4p}{(2\pi)^4}\,F(p+q/2)F(-p+q/2) \\
\end{array}
\right. \label{eq3}
\end{eqnarray}
Then equation (\ref{eqbethsalp}) takes the form:
\begin{eqnarray}
\Gamma_{12}(q)\chi_{12}(q)= 1+ \Gamma_{11}(q)\chi_{11}(q)
\label{eq5}
\end{eqnarray}

\begin{figure}\centering
\vbox to 60mm{\epsfysize=60mm \epsfbox{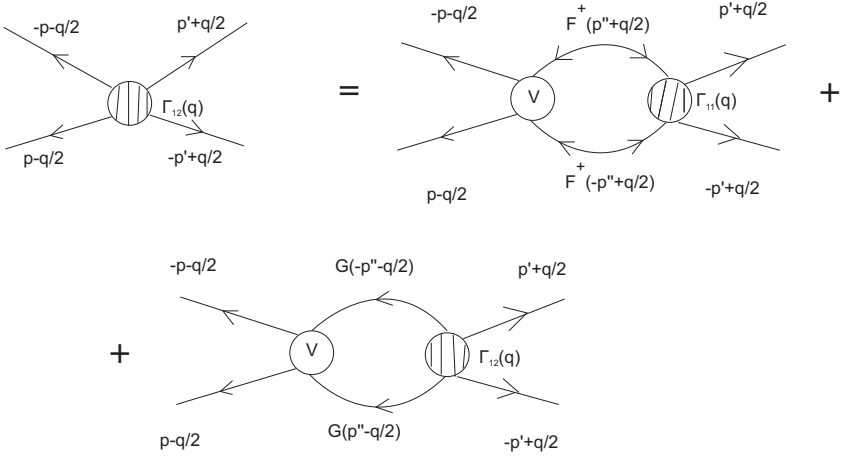}}
\caption{The Bethe-Salppeter equation for $\Gamma_{12}$}
\label{fig3bis}
\end{figure}

Now we have to  derive a second Bethe-Salpeter equation for
$\Gamma_{12}(q)$. Graphically it has the form shown on Fig.
\ref{fig3bis}. A small difference is that the anomalous Green's functions
$F^+(p)$ appearing in this equation have their arrow pointing outside, instead
of inside as in Fig. \ref{fig2bis}. However for real $\Delta$, they are simply related
to the preceding ones by $F^+(p)=F_S(-p)$.

In algebraic form it reads:
\begin{eqnarray}
\Gamma_{12}(q)=-g\, \Gamma_{12}(q) \int \!\frac{d^4p''}{(2\pi)^4}\,
G(p''-q/2)G(-p''-q/2)-g\,\Gamma_{11}(q) \int \! \frac{d^4p''}{(2\pi)^4}\,
F^+(p''+q/2)F^+(-p''+q/2)
\label{eq6}
\end{eqnarray}
Renormalization requires again substitution of $G
G$ by $G G-G_0G_0$ and the replacement of $V$
by  $g=4\pi a/m$. We see naturally appearing in this equation the
above response functions Eq.(\ref{eq5}), but for the 4-momentum 
$-q$, leading to the following form for this second equation:
\begin{eqnarray}
\Gamma_{12}(q)\chi_{11}(-q)=\Gamma_{11}(q)\chi_{12}(-q)
\label{eq10}
\end{eqnarray}
From Eq.(\ref{eq5}) and Eq.(\ref{eq10}) the vertices $\Gamma_{11}$
and $\Gamma_{12}$ are immediately obtained.
However we are only interested in the pole of these vertices, for which
they diverge. This allows us to solve only the homogeneous equations corresponding to 
Eq.(\ref{eq5}) and Eq.(\ref{eq10}), leading to the following equation for the mode:
\begin{eqnarray}\label{eqmodmax}
\chi_{11}(q)\chi_{11}(-q)=\chi^2_{12}(q)
\end{eqnarray}
where we have used the fact that, taking the explicit form for $F(k)$
in Eq.(\ref{eq3}), one has $\chi_{12}(-q)=\chi_{12}(q)$. This result has already been
derived in Ref. \cite{pps}.

In order to obtain a more convenient equation for the mode, we can perform
the integration over the frequency variable in Eq.(\ref{eq3}). This is easily done
by closing the contour in the upper half-plane where the quantity to be integrated
has two poles, located at $iE_+$ and $iE_-$. Here we have introduced the convenient
notation $E_{\pm}= E_{k \pm q/2}$, and we will use similarly $\xi_{\pm}= \xi_{k \pm q/2}$.
The results are:

\begin{eqnarray}
-\chi_{11}(\omega,{\bf q})=\frac{1}{2}\int\frac{d^3{\bf k}}{(2\pi)^3}
\left[\frac{(E_+ + E_-)(E_+E_- + \xi_+\xi_-) + i\omega(E_+\xi_- +
E_-\xi_+)}{E_+E_- \left[ (E_+ + E_-)^2 + \omega^2
\right]}- \frac{1}{E_k}\right]
 \label{eq18}
\end{eqnarray}
\begin{eqnarray}
\chi_{12}(\omega,{\bf q}) = \frac{1}{2} \int \frac{d^3
{\bf k}}{(2\pi)^3} \frac{\Delta^2}{E_+E_-} \frac{E_+ + E_-}{\left[
(E_+ + E_-)^2 + \omega^2 \right]}
\label{eq20}
\end{eqnarray}
where, in Eq.(\ref{eq18}), we have made used of the gap equation Eq.(\ref{eqgap}) to get
rid of the term $1/g$ in Eq.(\ref{eq3}). Finally we have to go back to ordinary frequencies by the "inverse" Wick transformation $i\omega \rightarrow \omega $ to recover the results of section \ref{disp} for the pole corresponding to the collective mode.

\subsubsection{KINETIC EQUATION}

This kinetic equation approach is the generalization of the well known Landau's kinetic equation for Fermi liquids to the superfluid phase described by BCS theory. It provides a dynamic self-consistent mean-field theory which is gauge invariant and satisfies from the start conservation laws. In particular, in the appropriate limit, it reduces to hydrodynamics as it should. It has been derived from the BCS formalism by Stephen \cite{stephen}, generalized by Betbeder-Matibet and Nozi\`eres \cite{bmn} to include Fermi liquid effects and later used by W\"olfle \cite{wolfl} in the case of superfluid $^3$He. Its application to cold gases has been pointed out by Randeria \cite{gss}. A specific application close to the present derivation where details can be found is Ref. \cite{rcjltp}. Here we use only the collisionless version of this equation.

To obtain easily this equation in a compact form it is convenient to write the standard BCS Hamiltonian, which rules the dynamics of the system:
\begin{eqnarray}
H_{BCS} =  \sum_{k\,\alpha } \xi^{}_{k} \;c^{\dag}_{k\,\alpha } c^{}_{k\,\alpha } 
+  \sum_{k} (\Delta  \, c^{\dag}_{k \uparrow} c^{\dag}_{-k \downarrow} + \rm{h.c.})
\end{eqnarray}
with Nambu spinor notations as (within an additive constant):
\begin{eqnarray}
H_{BCS}=\, \sum_{k \lambda \mu }
c^{\dag}_{k \lambda } \,^{tr}{\hat \epsilon}_{k \lambda \mu } \,c_{k \mu }
\end{eqnarray}
where $\lambda,\mu = 1$ or $2$, and by definition $c_{k,1}=c_{k\,\uparrow}$ and
$c_{k,2}=c^{\dag}_{-k\,\downarrow}$. This corresponds to a unitary transformation and implies the anticommutation relations $ [c_{k\, \lambda},c_{k'\,\mu }^{\dag}]=\delta_{k, k'}\,\delta_{\lambda\mu }$ and $ [c_{k\, \lambda},c_{k'\,\mu }]=0$. The 2x2 matrix ${\hat \epsilon}_{k \lambda \mu}$ is given by:
\begin{eqnarray}\label{eqhmatr}
{\hat \epsilon} _{k} = \left(\begin{array}{cc}\xi_k & \Delta^{*} \\\Delta & -\xi_k
\end{array}\right)
\end{eqnarray}
and $^{tr}$ indicates the transposition.

We consider now a small perturbation $\delta U$ with frequency $\omega $ and wavevector ${\bf q}$, coupled only to density fluctuations for simplicity. This gives rise to a corresponding fluctuating part $\delta \Delta $ of the order parameter, with the same frequency and wavevector. These perturbations give an additional contribution to the Hamiltonian:
\begin{eqnarray}
\delta H =\,\sum_{k \lambda \mu }
c_{k+q/2, \lambda }^{\dag} \,^{tr}\delta {\hat \epsilon}_{\lambda \mu} \,c_{k-q/2, \mu}
\end{eqnarray}
with $\delta {\hat \epsilon}_{11}=-\delta {\hat \epsilon}_{22}=\delta U $ and $\delta {\hat \epsilon}_{21}=(\delta {\hat \epsilon} _{12})^{*}=\delta\Delta$. The time evolution of the single particle density matrix:
\begin{eqnarray}
({\hat n}_{k\,k'})_{\lambda\mu }=\langle c_{k\,\lambda }^{\dag}c_{k'\,\mu } \rangle 
\end{eqnarray} 
can then be obtained from Heisenberg equations of motion and commutation relations. One finds that the fluctuating part of this density matrix $\delta {\hat n}_k = {\hat n}_{k-q/2,\,k+q/2}$ satisfies the following kinetic equation:
\begin{eqnarray}\label{eqkin}
\omega \delta {\hat n}_k =  \delta {\hat n}_k  \;{\hat \epsilon} ^{0}_{k+q/2} - {\hat \epsilon} ^{0}_{k-q/2} \; \delta {\hat n}_k
+ {\hat n}^{0}_{k-q/2} \;\delta {\hat \epsilon}_k  - \delta {\hat \epsilon}_k  \;{\hat n}^{0}_{k+q/2}
\end{eqnarray}
Here ${\hat \epsilon} ^{0}_{k}$ is the equilibrium value of the matrix ${\hat \epsilon}_{k}$ given by Eq.(\ref{eqhmatr}), $\Delta $ taking its equilibrium value, and ${\hat n}^{0}_{k}$ is related to the corresponding $T=0$ equilibrium value ${\hat n}^{0}_{k\,k'}$  of ${\hat n}_{k\,k'}$ given by ${\hat n}^{0}_{k\,k'}= {\hat n}^{0}_{k}\, \delta_{k\,k'}$, and it is given by:
\begin{eqnarray}
{\hat n}^{0}_{k} = \frac{1}{2}\;(1-\frac{{\hat \epsilon}^{0}_{k} }{E_{k}})
\end{eqnarray}
Naturally the order parameter is self-consistently related to the off-diagonal part of ${\hat n}_{k\,k'}$, both for the equilibrium as well as for the fluctuating part. These equations are conveniently combined to eliminate the interaction strength $V$ and yield the self-consistent relations:
\begin{eqnarray} \label{eqself}
\sum_{k}(\delta {\hat n}_k)_{21}+\frac{\delta \Delta }{2E_k} = 0 \hspace{10mm} \sum_{k}(\delta {\hat n}_k)_{12}+\frac{\delta \Delta^{*} }{2E_k} = 0
\end{eqnarray}
which allows to close the set of equations. In this way the response to the perturbation $\delta U$ can be calculated, and its pole provides the mode dispersion relation. In the present case, since we are only interested in this pole and not in the response function, we may set from the start $\delta U = 0$ and the equation for the mode dispersion relation will just be obtained from the compatibility condition for Eq.(\ref{eqself}.) The algebra is simplified if one premultiplies the matrix equation (\ref{eqkin}) by $U_{k-q/2}$ and postmultiplies it by $U_{k+q/2}$ where:
\begin{eqnarray}\label{}
U _{k} = \left(\begin{array}{cc}u_k & v_k \\\ -v_k & u_k \end{array}\right)
\end{eqnarray}
where $u_k=[(1+\xi_k/E_k)/2]^{1/2}$ and $v_k=[(1-\xi_k/E_k)/2]^{1/2}$ are the well-known coefficients of the Bogoliubov-Valatin transformation, which diagonalizes both ${\hat \epsilon} ^{0}_{k}$ and ${\hat n}^{0}_{k}$ . This leaves only diagonal elements for the matrix equation which is then easily solved. We skip this algebra and give only the result for the off-diagonal elements of $\delta {\hat n}_k(\omega ,{\bf q})$:
\begin{eqnarray}\label{eqdn}
(\delta {\hat n}_k)_{21}(\omega ,{\bf q})=(\delta {\hat n}_k)^{*}_{12}(-\omega ,{\bf q})=
\frac{1}{2}\,\frac{1}{\omega ^2-(E_++E_-)^2}
\left[\, \delta \Delta \left\{ \omega \,(\frac{\xi_-}{E_-}+\frac{\xi_+}{E_+}) + (E_++E_-)(1+\frac{\xi_-\xi_+}{E_-E_+}) \right\} \right. \\ \nonumber
-\left.\delta \Delta^{*}(E_++E_-)\frac{\Delta ^2}{E_-E_+}\, \right]
\end{eqnarray}
Intermediate steps can be found in Ref. \cite{rcjltp}. It is easy to check that this equation gives the proper result in the static and uniform limit $\omega =0,\,{\bf q}=0$. Finally it is worth noticing that Eq.(\ref{eqkin}) yields automatically the mass conservation equation, by taking the trace and summing over ${\bf k}$.

\subsubsection{RESPONSE FUNCTION}

Here we give for completeness the explicit expression of the density-density response function $\chi(q,\omega ) \equiv \delta n/\delta U$, where $\delta n =  \sum_{k} (\delta n_k)_{11}$ is the single spin state density fluctuation produced by the perturbation $\delta U$. In addition to the integrals introduced in section \ref{disp} we set:
\begin{eqnarray}\label{I}
{I}(\omega ,q)= \int_{0}^{\infty} \!\!{k}^{2}d{k} \int_{0}^{1}\!\!du\,\frac{({\xi}_{+}+{\xi}_{-})({E}_{+}+{E}_{-})}{{E}_{+}{E}_{-}[({E}_{+}+{E}_{-})^{2}-\omega ^{2}]}
\end{eqnarray}
\begin{eqnarray}\label{I'}
{I'}(\omega ,q)= \int_{0}^{\infty} \!\!{k}^{2}d{k} \int_{0}^{1}\!\!du\,\frac{{E}_{+}+{E}_{-}}{{E}_{+}{E}_{-}[({E}_{+}+{E}_{-})^{2}-\omega ^{2}]} = \frac{{I}_{11}(\omega ,q)-{I}_{22}(\omega ,q)}{2 \Delta ^{2}}
\end{eqnarray}
\begin{eqnarray}\label{I''}
{I''}(\omega ,q)= \int_{0}^{\infty} \!\!{k}^{2}d{k} \int_{0}^{1}\!\!du\,\frac{({E}_{+}+{E}_{-})}{{E}_{+}{E}_{-}}\, \frac{{E}_{+}{E}_{-}-{\xi}_{+}{\xi}_{-}+{\Delta}^{2}}{({E}_{+}+{E}_{-})^{2}-\omega ^{2}}
\end{eqnarray}
Then the response function is given by:
\begin{eqnarray}\label{chi}
\chi(q,\omega ) =-\frac{1}{4\pi ^2}\left[ I''(\omega ,q) - \Delta ^{2}\;\frac{I_{11}(\omega ,q)I^{2}(\omega ,q)+\omega ^{2}I_{22}(\omega ,q)I'^{2}(\omega ,q)-2\omega ^{2} I_{12}(\omega ,q)I (\omega ,q)I'(\omega ,q)}{I_{11}(\omega ,q)I_{22}(\omega ,q)-\omega ^{2}I_{12}^{2}(\omega ,q)}\right]
\end{eqnarray}
The first term, proportional to $I''$ is the simple and naive BCS result one would obtain by applying the standard linear response formula to the BCS ground state. Its poles exhibit, at low momenta,  the typical gap $2\Delta$ of BCS theory. The second term is the contribution coming from the dynamics of the order parameter, and plays a crucial role in providing the "phonon" dynamics absent in BCS theory. In particular this term restores the $f$-sum rule, which results from particle conservation and which is violated by simple BCS theory, and its poles yield naturally the dispersion relation Eq.(\ref{eqdisprel}) of the collective mode. Note that the imaginary part of the response function Eq.(\ref{chi}) is proportional to the dynamic structure factor whose behaviour along the BEC-BCS crossover has been recently investigated in \cite{cgs}.

\subsection{VICINITY OF THE CONTINUUM ON THE BCS WEAK COUPLING SIDE}

One can make, in the limit $\omega \rightarrow 2 \Delta $ where the collective mode is very near the continuum threshold, an analytical study of Eq.(\ref{eqdisprel}). In this limit the denominator $({E}_{+}+{E}_{-})^{2}-\omega ^{2}$ in the integrals ${I}_{11}$,${I}_{12}$ and ${I}_{22}$ is nearly zero when ${\xi}_{+}$ and ${\xi}_{-}$ goes to zero. This condition on ${\xi}_{+}$ and ${\xi}_{-}$ is equivalent to require that $u$ is small and ${k}$ is near ${k}_{th} = \sqrt{2m\mu -q^2/4}$. However a closer examination shows that the numerators in ${I}_{12}$ and ${I}_{22}$ go also to zero in this case, and as a result these two integrals have actually a finite limit when $\omega \rightarrow 2 {\Delta}  $. On the other hand there is indeed a divergence which appears in ${I}_{11}$. Since, from Eq.(\ref{eqdisprel}), ${I}_{11}$ has to be finite when the dispersion relation is satisfied, this divergence has to be cancelled in some way. Taking for example Eq.(\ref{I11p}) for ${I}_{11}$, one sees that only the $\omega^2$ term has a divergence. It is easily seen that, within a constant prefactor, the divergent term of ${I}_{11}$ behaves as $(1/q) \ln[1/(2 {\Delta} - \omega )]$. In order to compensate for the logarithmic divergence, $q$ has to grow like $\ln[1/(2 {\Delta} - \omega )]$ when $\omega \rightarrow 2 {\Delta}$. This means that the dispersion relation $\omega(q)$ approaches the limit $\omega = 2 \Delta$ with an exponential behaviour. This result is completely consistent with numerical results. 

Naturally this result holds only as long as ${k}_{th}$ exists. This implies as a necessary condition that $q < 2 \sqrt{2m\mu }$. One can consider directly the case $q = 2 \sqrt{2m\mu }$. It is easily seen that the qualitative situation is not changed, i.e. ${I}_{11}$ is divergent for $\omega = 2 {\Delta}$ whereas the other integrals are finite. This implies again that one must have $\omega <  2 {\Delta}$ in order to satisfy the mode dispersion relation. On the other hand the quantitative situation is slightly different because the divergence is produced by contributions to the integral coming from the vicinity of ${k}=0$. Then, when one goes in the domain $q > 2 \sqrt{2m\mu }$, the situation changes qualitatively because as we have seen in section \ref{thresh} the minimum of ${E}_{+}+{E}_{-}$ does not correspond anymore to the condition ${\xi}_{+}={\xi}_{-}=0$, but rather to ${\xi}_{+}={\xi}_{-}=q^2/(8m) - \mu $, obtained for ${k}=0$. Because of the phase space factor ${k}^2$ in all the integrals, the three integrals ${I}$ are finite at the continuum threshold. As it is seen in section \ref{contact}, the collective mode does indeed merge in the continuum in the weak coupling regime, but this occurs always for $q > 2 \sqrt{2m\mu }$. However when one goes toward weaker coupling, this merging point is increasingly close to the $q = 2 \sqrt{2m\mu }$ point.

\subsection{FIRST ORDER EXPANSION ON THE BEC SIDE}

We want to expand the dispersion relation Eq.(\ref{eqdisprel}) on the BEC side to first order in ${(\Delta }/|\mu |)^2$. This means going one step further than what we have done in the preceding calculation in section \ref{bec}. This turns out to be not so complicated analytically. Indeed the only different term between ${I}_{11}$ and ${I}_{22}$, namely the ${\Delta }^2$ term in Eq.(\ref{I11}) and (\ref{I22}), gives a negligible ${\Delta }^4$ contribution in the product ${I}_{11}{I}_{22}$ coming in the dispersion relation Eq.(\ref{eqdisprel}). This implies that we can omit this ${\Delta }^2$ term in Eq.(\ref{I11}) and (\ref{I22}) and take ${I}_{11}={I}_{22}$. Hence the dispersion relation simplifies again into ${I}_{11}+\omega {I}_{12}=0$ as precedingly. Moreover we have to first order in ${\Delta }^2$, ${E}_{\pm}={\xi}_{\pm} + {\Delta }^2/2{\xi}_{\pm}$. This implies that we have at this order, for the numerators coming in ${I}_{11}$ and ${I}_{12}$, ${E}_{+}{\xi}_{-}+{\xi}_{+}{E}_{-}={E}_{+}{E}_{-}+{\xi}_{+}{\xi}_{-}$, so that the dispersion relation becomes:
\begin{eqnarray}\label{eqdisprel1}
\int_{0}^{\infty} \!\!{k}^{2}d{k} \left[ \int_{0}^{1}\!\!du\,(1+\frac{{\xi}_{+}{\xi}_{-}}{{E}_{+}{E}_{-}}) \,\frac{1}{{E}_{+}+{E}_{-}-2\omega } - \frac{1}{{E}}\,\right ] = 0
\end{eqnarray}
Here we use reduced units where all the energies are expressed in units of $|\mu |$ and all wavevectors in units of $k_0 \equiv \sqrt{2m|\mu |}$, with the notations $\Omega = \omega /|\mu |$ and $Q=q/k_0$. All the quantities in reduced units are indicated by a bar over the corresponding symbol. In terms of these reduced units, a first order expansion leads to:
\begin{eqnarray}\label{eqdisprel1}
\frac{\pi }{8}(\Omega - \frac{Q^2}{2})=\frac{{\bar \Delta }^2}{4}\,\int_{0}^{\infty} \!\!d{\bar k}\; \frac{{\bar k}^{2}}{1+{\bar k}^{2}}\int_{0}^{1}\!\!du\,\left[ (\frac{1}{{\bar \xi}_{+}^2}+\frac{1}{{\bar \xi}_{-}^2})+\frac{1}{1+{\bar k}^{2}}(\frac{1}{{\bar \xi}_{+}}+\frac{1}{{\bar \xi}_{-}})-\frac{2}{(1+{\bar k}^{2})^2} \right ] 
\end{eqnarray}
After performing the integrations one obtains:
\begin{eqnarray}\label{eqdispdev}
\Omega - \frac{Q^2}{2} = \frac{{\bar \Delta }^2}{4}\,\frac{16-Q^2}{16+Q^2}
\end{eqnarray}
equivalent to:
\begin{eqnarray}\label{eqdispdev1}
\Omega^2 =  Q^2\;\frac{{\bar \Delta }^2}{4}\,\frac{16-Q^2}{16+Q^2}+\left( \frac{Q^2}{2} \right)^2
\end{eqnarray}
up to first order in ${\bar \Delta }^2$.


\begin{references}
\bibitem{note1}This is a standard, but rough and somewhat misleading statement. Actually on the BEC side only the full limit of infinitely diluted molecules is properly described, but the interaction between molecules is known to be quantitatively incorrect since the actual molecule-molecule scattering length, calculated by solving the 4-body problem, turns out to be $0.6\,a$ \cite{petrov,bkkcl} instead of the $2\,a$ result given by the BCS ansatz. Similarly on the BCS side the critical temperature is not given by a simple BCS formula with bare interaction, but has to take into account indirect interactions \cite{gmb} which lead to a modification of the numerical coefficient coming in the standard formula.
\bibitem{note2}If a diagonal mean-field is included in the most straightworward way, it would lead on the BCS side to a collapse instability under the effect of attractive forces which is not observed experimentally.
\bibitem{note}In earlier experiments \cite{collmode} the agreement was not so good, and it had been suggested \cite{rcxlcomm} that this might be due to the fairly high frequency of the radial oscillation.
\end{references}
\end{document}